\documentstyle[prl,twocolumn,aps]{revtex}
\input psfig
\begin{document}
\draft

 \newcommand{\mytitle}[1]{
 \twocolumn[\hsize\textwidth\columnwidth\hsize
 \csname@twocolumnfalse\endcsname #1 \vspace{1mm}]}

\mytitle{
\title{Real-time renormalization group and charge fluctuations
in quantum dots}

\author{Herbert Schoeller$^{1,2}$ and J\"urgen K\"onig$^{3}$}

\address{
$^1$Forschungszentrum Karlsruhe, Institut f\"ur 
Nanotechnologie, 76021 Karlsruhe, Germany\\
$^2$Institut f\"ur Theoretische Festk\"orperphysik, Universit\"at
Karlsruhe, 76128 Karlsruhe, Germany\\
$^3$Department of Physics, Indiana University, Bloomington,
Indiana 47405, USA}

\date{\today}

\maketitle

\begin{abstract}
We develop a perturbative renormalization-group method in real time to 
describe nonequilibrium properties of discrete quantum systems coupled 
linearly to an environment. 
We include energy broadening and dissipation and develop a cutoff-independent 
formalism. 
We present quantitatively reliable results for the linear and nonlinear 
conductance in 
the mixed-valence and empty-orbital regime of the nonequilibrium Anderson 
impurity model with finite on-site Coulomb repulsion.
\end{abstract}
}

Renormalization group (RG) methods are standard tools to describe various 
aspects of condensed-matter problems beyond perturbation theory
\cite{anderson-wilson}.
Many impurity problems have been treated by numerical RG with excellent 
results both for thermodynamic quantities and spectral densities 
\cite{nrg,costi}.
These RG techniques, however, cannot describe nonequilibrium properties like
the nonlinear conductance, the nonequilibrium stationary state, or the full 
time development of an initially out-of-equilibrium system.
To address these aspects we present here a perturbative RG method, formulated 
for strongly-correlated quantum systems with a finite number of states coupled 
linearly to external heat or particle reservoirs.
Fundamentally new, we work on a Keldysh contour and generate non-Hamiltonian 
dynamics during RG, which captures the physics of finite life times and 
dissipation. 
Furthermore, no initial or final cutoff in energy or time space is needed, 
i.e., large and small energy scales are accounted for correctly like in 
flow-equation methods \cite{flow equation}. 
Although correlation functions can also be studied, physical quantities like 
spin and charge susceptibilities or the current can be calculated directly 
without the need of nonequilibrium Green's functions. 

The purpose of our RG technique is to describe quantum fluctuations which are 
induced by strong coupling between a small quantum system and an environment. 
There are several recent experiments which show the importance of quantum 
fluctuations in metallic single-electron transistors \cite{joy-etal} and 
semiconductor quantum dots \cite{gol-etal,cro-etal,kondo-exp} 
(see \cite{schoeller} for an overview over theoretical papers). 
Due to the renormalization of resistance and local energy excitations, 
anomalous line shapes of the conductance have been observed, which can not be 
explained by golden-rule theories. 
Therefore, the RG approach presented here can treat simultaneously strong
coupling to the reservoirs, strong Coulomb interaction on the island, and 
finite bias voltage. 
We apply the technique to a quantum dot with one spin-degenerate state and 
present for the first time quantitatively reliable results for the nonlinear 
conductance in the mixed-valence and empty-orbital regime. 
For applications to the single-electron box and the 
single-electron transistor we refer to Ref.~\cite{rg-set}.

We consider the nonequilibrium Anderson impurity model with finite on-site 
Coulomb repulsion $U$.
It consists of a single, spin-degenerate state in a quantum dot which is
coupled via tunneling to a left and right reservoir.
The Hamiltonian $H=H_{res}+H_D+H_T$ consists of three parts. 
$H_{res}=\sum_{k\sigma r}\epsilon_{k\sigma r}a^\dagger_{k\sigma r} 
a_{k\sigma r}$ describes two reservoirs, $r=L,R$, where $k$ labels the states 
and $\sigma$ denotes the spin. The isolated quantum dot is given by 
$H_D=\sum_\sigma \epsilon_\sigma n_\sigma + U n_\uparrow n_\downarrow$ 
with $n_\sigma=c^\dagger_\sigma c_\sigma$. 
The coupling to the reservoirs is described by the standard tunneling 
Hamiltonian 
$H_T=\sum_{k\sigma r}(T^r_k a^\dagger_{k\sigma r} c_\sigma\,+\,h.c.)$. 
The goal is to evaluate the reduced density matrix of the dot system, 
$\rho_D(t)={\rm Tr}_{res} \rho(t)$, and the expectation value of the current 
operator $I=I_L= ie\sum_{k\sigma} 
(T^L_k a^\dagger_{k\sigma L} c_\sigma\,-\,h.c.)$.

We start with the time evolution of the dot system,
$\rho_D(t) ={\rm Tr}_{res} \left\{e^{-iHt}\rho(0)e^{iHt} \right\}$.
Initially, the system is decoupled, and each reservoir is in thermal 
equilibrium, $\rho(0)=\rho_D(0)\rho^{eq}_L\rho^{eq}_R$.
Nonequilibrium situations arise when the two reservoirs have different 
electrochemical potentials $\mu_r$. 
As in Refs.~\cite{schoeller,schoeller-koenig-schoen} we expand the
forward/backward propagators $\exp({\mp iHt})$ in $H_T$,
and perform the trace ${\rm Tr}_{res}$ by applying
Wick's theorem with respect to the reservoir field operators.
All terms can be represented diagrammatically as shown in
Fig.~\ref{fig-keldysh}. Tunneling vertices are ordered along a closed Keldysh 
contour. They are connected in pairs by the contractions (dashed lines in 
Fig.~\ref{fig-keldysh})
$\gamma^+_r(t)=\sum_k |T^r_k|^2{} 
\langle a^\dagger_{k\sigma r}(t)a_{k\sigma r}\rangle$
or $\gamma^-_r(t)=\sum_k |T^r_k|^2 
\langle a_{k\sigma r}(t)a^\dagger_{k\sigma r}\rangle$,
depending on the relative time-ordering of the vertices on the
contour. We obtain for $r=L,R$,
\begin{equation}
\label{contraction}
   \gamma^\pm_r(t)={-i\Gamma_r e^{\pm i \mu_r t}
       \over 2 \beta\sinh[\pi(t-i0^+)/\beta]}
\end{equation}
with $\beta=1/(k_B T)$, 
and $\Gamma_r=2\pi\sum_k |T^r_k|^2\delta(E-\epsilon_{k\sigma r})$.
The solid line in Fig.~\ref{fig-keldysh} represents free time evolution
of the dot. As a result, we have obtained an effective theory of the dot 
degrees of freedom while the reservoirs have been integrated out.

To derive a kinetic equation we will call diagrams irreducible if any vertical
cut crosses at least one dashed line. Fig.~\ref{fig-keldysh} 
shows four such irreducible blocks (a)-(d).
We denote the sum over all irreducible diagrams between $t'$ and $t$ by 
the kernel $\Sigma(t-t')$.
The full time evolution, i.e., a sequences of $\Sigma$ blocks, can be 
described by a self-consistent equation. 
Differentiating with respect to time $t$ leads to the standard
kinetic equation \cite{schoeller,schoeller-koenig-schoen} 
\begin{equation}
\label{kinetic equation}
   \dot{\rho}_D(t) + iL_D \rho_D(t) = \int^t_0 dt' \, 
\Sigma(t-t') \rho_D(t')\,\,.
\end{equation}
Here, $L_D=[H_D,\cdot]$ and $\Sigma$ are superoperators acting 
on $\rho_D(t)$, i.e., the $4^4=256$ matrix elements 
$\Sigma_{s_1 s_1^\prime,s_2 s_2^\prime}$ are labeled by
four dot-states with $s_{1/2}$ ($s_{1/2}^\prime$) referring to the
forward (backward) propagator.
The l.h.s of Eq.~(\ref{kinetic equation}) describes the time evolution of
the dot in the absence of tunneling whereas the r.h.s contains the dissipative 
part which drives the dot distribution into a stationary state. 
In Laplace space Eq.~(\ref{kinetic equation}) reads
$[z-L_D-i\tilde{\Sigma}(z)]\tilde{\rho}_D(z)=i\rho_D(0)$. 
Thus, the knowledge of $\tilde{\Sigma}(z)$ provides
the full time evolution of the dot distribution.
The stationary solution follows from 
$[L_D+i\tilde{\Sigma}(0)]\rho_D^{st}=0$.
The different electrochemical potentials of the reservoirs
(leading to a nonequilibrium stationary state) enter via the
pair contractions (\ref{contraction}) in the kernel $\tilde{\Sigma}(z)$.

{\it Sequential tunneling} or the {\it golden rule} results
\cite{averin+beenakker} can be recovered by calculating the kernel 
$\tilde{\Sigma}(z)$ perturbatively in first order in the tunneling coupling 
$\Gamma_r$.
The aim of the present paper, however, is to go beyond and calculate the 
kernel nonperturbatively by a systematic RG procedure. 
The idea is to integrate out all contraction lines one after another, beginning
with the shortest one.
Fig.~\ref{fig-keldysh} shows four examples (a)-(d) in second order in $\Gamma$.
In these examples, the lines with index $R$ are integrated out first.
This leads to a renormalization of the dot propagator in (a) and the tunneling 
vertex in (c).
In (b) and (d) the contraction line $R$ connects the forward with the backward
propagator. 
This element does not occur in any equilibrium theory but arises for a Keldysh 
contour in a natural way. 
Such contractions lead to non-Hamiltonian dynamics for the dot distribution 
and, therefore, account for dissipation.
To treat these cases in the same way as (a) and (c), it is convenient to 
view the forward and backward propagator formally as one double line, see 
Fig.~\ref{fig-liouville}.
As a consequence, (b) and (d) fall into the same topological classes as (a) 
and (c), respectively.
The price is that a ''state'' on the double line has to be specified by two 
dot states, one for the upper and one for the lower propagator. This leads to 
the superoperator matrix notation used in Eq.~(\ref{kinetic equation}).
The tunneling vertices $C^p_\mu$ on the double line act on these 
double states.
There are 16 different vertices, specified by four indices, $p$ and 
$\mu\equiv \eta\sigma r$, where $p=\pm$ indicates whether the unrenormalized 
vertex acts on the forward/backward propagator, $\eta=\pm$ represents 
tunneling in/out, $\sigma=\uparrow,\downarrow$ is the spin, and $r=L,R$ refers 
to the reservoir index.
As shown in example (d) of Fig.~\ref{fig-keldysh}, after being renormalized a
vertex will in general act on both the forward and backward propagator 
simultaneously. 
Double lines without vertices corresponds to a free time evolution 
$\exp({-iL_Dt})$ of the dot (with renormalized $L_D$).
Finally, we denote the rightmost (leftmost) vertex of the kernel 
$\tilde{\Sigma}$ by $A^p_\mu$ ($B^p_\mu$).
They are renormalized in a different way than $C^p_\mu$.

The flow parameter $t_c$ in our RG procedure is the largest length of those 
contractions that have already been integrated out.
An infinitesimal RG step is established by integrating out all contractions 
with a length $t$ between $t_c$ and $t_c+dt_c$. 
For $t_c\rightarrow\infty$ all contractions are
integrated out. We find the RG equations (summation over
repeated indices implicitly assumed)
\begin{eqnarray}
\label{rg-equations}
   {d\tilde{\Sigma}\over dt_c}&=&-\gamma^{pp'}_{\mu\mu'}(t_c)
   I_{pp'} A^p_\mu(t_c)B^{p'}_{\mu'}\label{rg-sigma}\,\,,\\
   {dL_D\over dt_c}&=&-i\gamma^{pp'}_{\mu\mu'}(t_c)
   I_{pp'} C^p_\mu(t_c)C^{p'}_{\mu'}\label{rg-L0}\,\,,\\
   {dC^p_\mu\over dt_c}&=&
   \gamma^{p_1p^\prime_1}_{\mu_1\mu^\prime_1}(t_c)
   \int_0^{t_c} dt \,\left[pp^\prime_1
   I_{p_1p^\prime_1} C^{p_1}_{\mu_1}(t)
   C^p_\mu\,\right. \nonumber \\
   &&  \hspace{1cm} \left. - \,\,C^p_\mu I_{p_1p^\prime_1} 
   C^{p_1}_{\mu_1}(t)\,\right]\,
   C^{p_1^\prime}_{\mu_1^\prime}(t-t_c)\label{rg-C}\,\,,\\
   {dA^p_\mu\over dt_c}&=&\gamma^{p_1p^\prime_1}_{\mu_1\mu^\prime_1}(t_c)
   \int_0^{t_c} dt \,\left[\,pp^\prime_1
   I_{p_1p^\prime_1} A^{p_1}_{\mu_1}(t)
   C^p_\mu \right. \nonumber \\
   &&  \hspace{1cm} \left. - \,\,A^p_\mu I_{p_1p^\prime_1} 
   C^{p_1}_{\mu_1}(t)\,\right]\,
   C^{p_1^\prime}_{\mu_1^\prime}(t-t_c)\label{rg-A}\,\,,\\
   {dB^p_\mu\over dt_c}&=&\gamma^{p_1p^\prime_1}_{\mu_1\mu^\prime_1}(t_c)
   \int_0^{t_c} dt \nonumber \\
   && \hspace{1cm}pp^\prime_1
   I_{p_1p^\prime_1} C^{p_1}_{\mu_1}(t)
   C^p_\mu B^{p_1^\prime}_{\mu_1^\prime}(t-t_c)\label{rg-B}\,\,,
\end{eqnarray}
We introduced $\gamma^{pp^\prime}_{\mu\mu^\prime}(t)=
\gamma^{-p^\prime\eta}_r(p^\prime t)\delta_{\bar{\mu}\mu^\prime}$
for the contraction between two tunneling vertices on the double line 
($\mu\equiv \eta\sigma r$, $\bar{\mu}\equiv -\eta\sigma r$,
$\mu^\prime\equiv \eta^\prime\sigma^\prime r^\prime$) to get a compact 
notation.
The interaction picture is defined by 
$C^p_\mu(t)=e^{iL_Dt}C^p_\mu e^{-iL_Dt}$,
$A^p_\mu(t)=e^{izt}A^p_\mu e^{-iL_Dt}$, and
$B^p_\mu(t)=e^{iL_Dt}B^p_\mu e^{-izt}$. The diagonal superoperator
$I_{pp^\prime}$ together with the factors $pp_1^\prime$ account for
possible minus signs due to Fermi statistics. We get
$(I_{pp^\prime})_{ss^\prime,ss^\prime}=(pp^\prime)^{N_s-N_{s^\prime}}$,
where $N_s$ is the particle number for state $s$.
All terms arise naturally from our intuitive picture set up in
Fig.~\ref{fig-liouville} except for the second term on the r.h.s of
Eqs.~(\ref{rg-C}) and (\ref{rg-A}). 
The latter are correction terms to account for correct time ordering\cite{com}.

To evaluate the average current we repeat the above derivation of the RG 
equations but replace the unrenormalized boundary vertex operator $A^p_\mu$ by 
the current vertex $I$.
The analog of Eq.~(\ref{rg-sigma}) gives the current kernel 
$\tilde{\Sigma}_{I}(z)$ and the current follows from 
$\langle I \rangle(z)={\rm Tr}_D \tilde{\Sigma}_{I}(z)\tilde{p}(z)$.
 
Taking matrix elements, all integrals in Eqs.~(\ref{rg-C})-(\ref{rg-B}) can be
calculated analytically, and pure differential equations are left which we 
solve numerically.
Most of the matrix elements which are initially zero remain unchanged under the
RG flow.
We end up with 26 equations for $\tilde \Sigma$ and $L_D$ each, and 192 
equations for $C$, $A$, and $B$ each to solve.

The RG flow of conventional poor man scaling and operator product expansion 
methods stops at some ''characteristic'' time (or inverse energy) scale.
All results, therefore, depend on a low-energy cutoff.
This is {\it not} the case in our formulation since we have {\it not} expanded 
the propagation $\exp(\pm iL_D t_c)$ in $t_c$ (it occurs in the interaction 
picture of the vertex operators) and can, therefore, integrate out {\it all} 
time scales. 
Furthermore, we work on a Keldysh contour and find that the renormalized 
$L_D$-operator of the dot system can no longer be represented as the 
commutator with a renormalized Hamiltonian $L_D\ne [H_D,\cdot]$. 
This corresponds to non-Hamiltonian dynamics and is needed to describe the
physics of dissipation and finite life times.

An important question concerns the validity range of the present
approach. Although no general statement is possible, the RG flow
itself will indicate the breakdown of the procedure by 
instabilities or increasing coupling constants.
Due to the neglect of higher-order vertex corrections, one
might not trust the procedure for too large coupling 
constants. However, for the specific example of the Anderson model in the 
mixed-valence regime (see below) we reproduce exact Bethe ansatz solutions 
within high accuracy even in the strong-coupling regime. 
The same was achieved for the ground-state energy of the single-electron box 
\cite{rg-set}. A possible explanation is the generation of rates
and energy broadening during RG which improves the accuracy and
sets another energy scale for the cutoff of the RG flow 
(for the present problem given by $\Gamma$). This is 
consistent with our numerical solution and with a recent 
analysis of poor man scaling on a Keldysh contour 
in Ref.~\cite{kaminski}.

In Fig.~\ref{fig-gatevoltage} we show the linear conductance $G$ 
for the Anderson model as function of $\epsilon=\epsilon_\sigma$, which 
is varied experimentally by the gate voltage. 
For $T=0$ and $\epsilon>-0.4\Gamma $, the deviation from the exact Friedel 
sum rule $G|_{T=0}=(2 e^2/h) \sin^2(\pi \langle n\rangle/2)$ is 
less than $2\%$. Here, $\Gamma/2=\Gamma_L=\Gamma_R$ and 
$\langle n\rangle =\sum_\sigma \langle n_\sigma \rangle$. 
In the same regime, the average occupation $\langle n\rangle$ agrees with
the exact Bethe ansatz solution \cite{tve-wie} within $3\%$ 
(inset of Fig.~\ref{fig-gatevoltage}). 
Since the RG flow is strongest if both temperature and bias voltage vanish, 
this comparism gives good support that our results are reliable in the whole 
mixed-valence and empty-orbital regime for all temperatures and bias voltages. 
Furthermore, we find excellent agreement with the exact Bethe ansatz solution 
for the magnetic and charge susceptibility at $\epsilon=0$ as function of
$T$ and $U$, respectively \cite{bethe}. 
Thus, the neglect of higher-order vertex terms seems to be justified. 

In the Kondo regime, i.e., for $\epsilon\lesssim -0.4\Gamma$, spin 
fluctuations are dominant and double-vertex terms become important.
In this regime, deviations from the Friedel sum rule 
can be seen in Fig.~\ref{fig-gatevoltage}. Experimentally, such 
deviations occur as well \cite{gol-etal,cro-etal,kondo-exp}, 
since it is difficult to reach the $T=0$ limit. 
Our theory accounts for charge fluctuations very well which are responsible for
the temperature-dependent renormalization of the dot level as shown by the 
shift of the maximum peak in Fig.~\ref{fig-gatevoltage}.
The fact that the conductance increases along the zero temperature result
on the r.h.s of the conductance peak is in good agreement with experiments 
\cite{gol-etal,cro-etal}.
A fit of the temperature dependence of the linear conductance to the 
experiment in Ref.~\cite{gol-etal} is shown
in Fig.~\ref{fig-temperature} for various values of $\epsilon$. 
The agreement is quite well except for 
$T\gtrsim 0.5\Gamma$ where the experimental conductance is higher. 
A reason for this discrepancy might be the presence of other levels with a 
level spacing of order $\Gamma$.
For a low-lying level the conductance increases monotonically with decreasing 
temperature and finally saturates. 
The saturated value at $T=0$ is not identical to the unitary limit 
$(e^2/h) 8\Gamma_L\Gamma_R/\Gamma^2$ since we are still in the mixed-valence 
regime.
However, the value agrees perfectly with the Friedel sum rule as already 
stated above.
In the empty-orbital regime, $\epsilon\gtrsim 0$, the conductance
shows a local maximum by varying $T$.
Here, the renormalized $\epsilon$ is above the Fermi level and will be 
occupied at high enough temperature. 
This increases the conductance with increasing $T$.

Fig.~\ref{fig-biasvoltage} shows $G=dI/dV$ as a function of 
the bias voltage for $\epsilon=-0.4\Gamma$. 
We obtain a zero-bias maximum with an amplitude increasing monotonically with 
decreasing $T$ (see Fig.~\ref{fig-temperature}). 
This maximum is {\it not} due to the presence of a Kondo resonance but stems 
from the renormalized dot level being near the Fermi levels of the 
reservoirs. 
The width of the peak in Fig.~\ref{fig-biasvoltage} scales with $\Gamma$ 
which is the relevant energy scale in the mixed-valence regime. 
Since the zero-bias maximum has a symmetric line shape, our results 
can be used for a fit of the important energy scale $\Gamma$ in 
experiments.

In summary we have presented a new RG technique to calculate nonequilibrium 
properties of dissipative quantum systems beyond perturbation theory. 
The method is very flexible and was applied to the Anderson-impurity model 
with finite $U$.
In the mixed-valence and empty-orbital regime the Friedel sum rule and
the exact Bethe ansatz solution were reproduced at zero temperature. 
The results are consistent with experiments. 

We acknowledge useful discussions with T. Costi, 
J.v. Delft, D. Goldhaber-Gordon, H. Grabert, M. Hettler, 
G. Sch\"on, K. Sch\"onhammer, and P. W\"olfle. We thank
U. Gerland for providing us with the Bethe ansatz results,
and D. Goldhaber-Gordon for the experimental data.
This work was supported by the Swiss National Science Foundation (H.S.) and 
by the ''Deutsche Forschungsgemeinschaft'' as part of ''SFB 195''.

\begin{figure}
\centerline{\psfig{figure=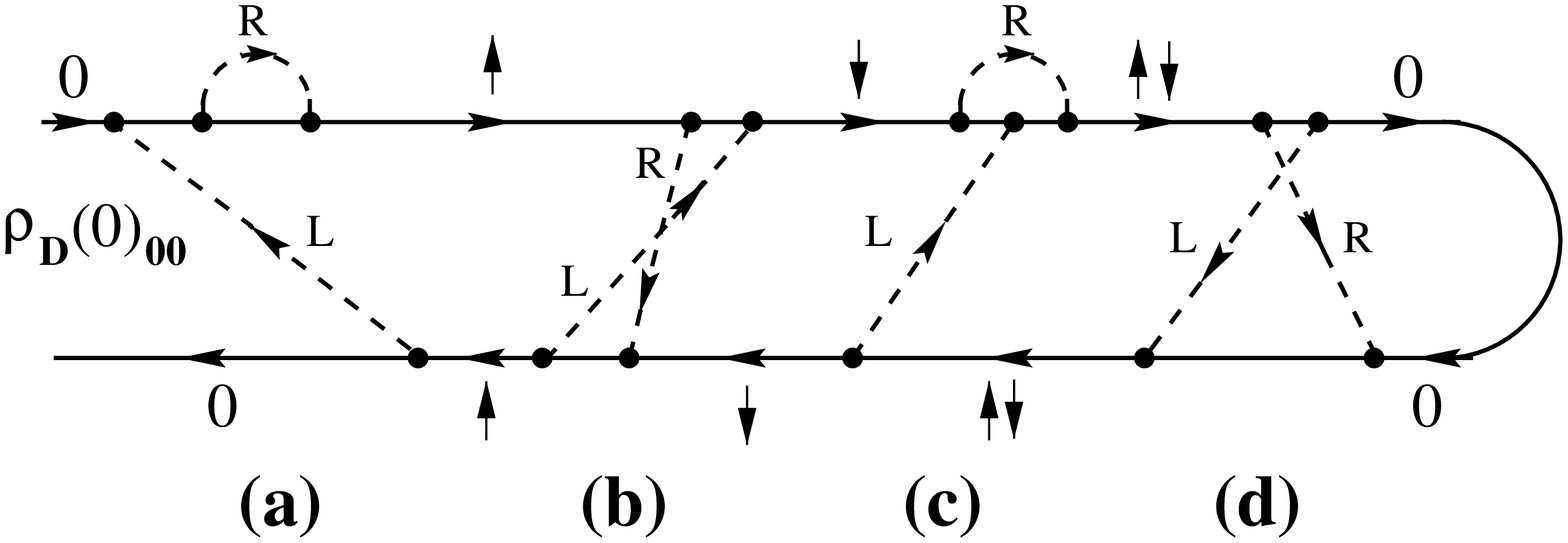,height=3cm,width=8cm}}
\vspace{.3cm}
\caption{Example of a diagram on the Keldysh contour for the 
matrix element $\rho_D(t)_{00}$. The contractions are labeled by the
reservoir indices $L$ and $R$.  }
\label{fig-keldysh}
\end{figure}

\begin{figure}
\centerline{\psfig{figure=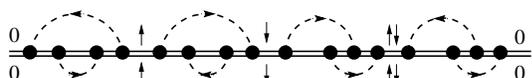,height=1.cm,width=7cm}}
\caption{The same figure as Fig.~\ref{fig-keldysh} but the two lines
taken together.}
\label{fig-liouville}
\end{figure}

\begin{figure}
\centerline{\psfig{figure=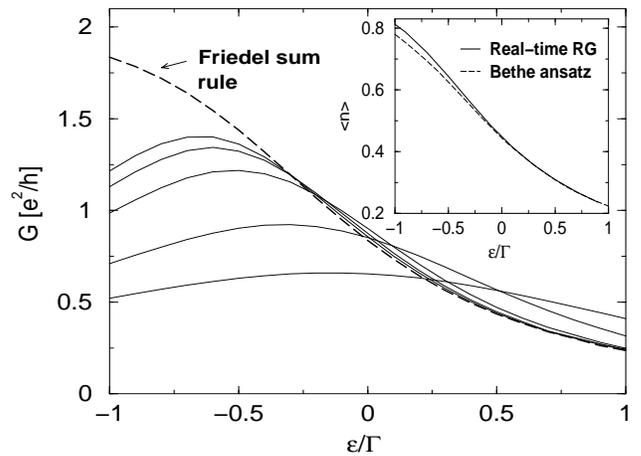,height=6.2cm}}
\caption{Linear conductance as a function of the dot level.
$T/\Gamma=0,0.05,0.1,0.25,0.5$ (solid lines from top to bottom) and 
$U=6\Gamma$. The inset shows the average occupation for $T=0$.}
\label{fig-gatevoltage}
\end{figure}

\begin{figure}
\centerline{\psfig{figure=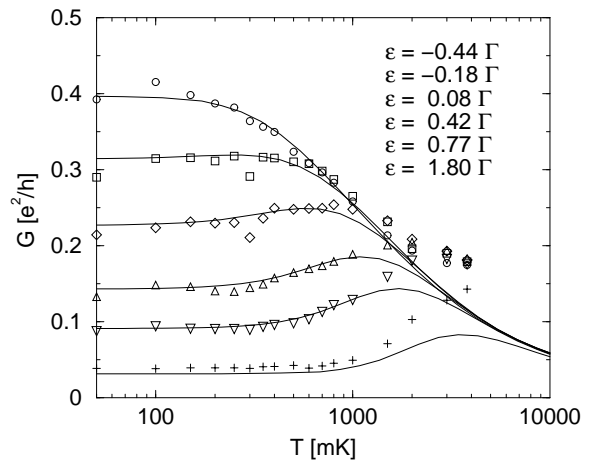,height=6.2cm}}
\caption[a]{Linear conductance as a function of temperature
in comparism to experiment \cite{gol-etal}. $U=6.2\Gamma$,
$8\Gamma_L \Gamma_R/\Gamma^2 = 0.6$, and $\Gamma=3423$mK.}
\label{fig-temperature}
\end{figure}

\begin{figure}
\centerline{\psfig{figure=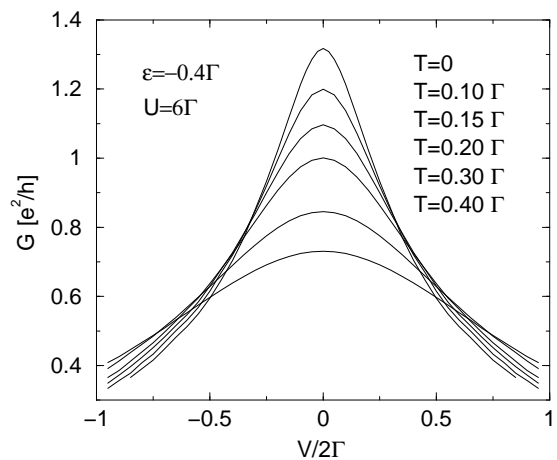,height=6.2cm}}
\caption{Nonlinear conductance as a function of the 
bias voltage.}
\label{fig-biasvoltage}
\end{figure}


\begin{references}
\bibitem{com}
Time misordering happens since the contractions in Fig.~\ref{fig-keldysh}a and 
\ref{fig-keldysh}b renormalizing $L_D$ are two-time objects but $L_D$ is 
treated after renormalization as a single-time one.
For further details of this subtle point and a more detailed derivation of the 
RG equations we refer to a forthcoming longer paper.
\end{references}
\end{document}